\documentclass[a4paper,fleqn,useAMS,usenatbib]{mnras}

\usepackage{psfrag,color,epsfig}

\usepackage{graphicx}	
\usepackage{amsmath}	
\usepackage{amssymb}	
\usepackage{multicol}	
\usepackage{bm}		
\usepackage{pdflscape}	
\usepackage{multirow}
\usepackage{tabularx}  
\usepackage{caption}

\usepackage[T1]{fontenc}
\usepackage{ae,aecompl}

\def\HI{\ion{H}{I}~}
\def\Hi{\ion{H}{I}}

\def\xHI{x_{\rm \ion{H}{I}}}
\def\xb{\bar{x}_{\rm \ion{H}{I}}}
\def\Tb{{T_{\rm b}}}
\def\tTb{\tilde{T}_{\rm b}}
\def\TTb{\tilde{T}_{\rm b2}}
\def\k{{\bm{k}}}

\def\r{{\bm{r}}}
\def\U{{\bm{U}}}
\def\thetavec{{\bm{\theta}}}
\def\cl{{\mathcal C}_{\ell}}
\def\n{\hat{\bm{n}}}
\def\v{\bm{v}}

\title[The light-cone EoR 21-cm signal]{Towards simulating and
  quantifying the light-cone EoR 21-cm signal}

\author[Mondal, Bharadwaj {\rm \&} Datta]{Rajesh Mondal,$^{1,2,3}$
  \thanks{E-mail:
    \href{mailto:rm@phy.iitkgp.ernet.in}{rm@phy.iitkgp.ernet.in}}  
Somnath Bharadwaj$^{1,2}$ and  Kanan K. Datta$^4$\\
$^1$ Department of Physics, Indian Institute of Technology Kharagpur,
Kharagpur -- 721302, India\\ 
$^2$ Centre for Theoretical Studies, Indian Institute of Technology
Kharagpur, Kharagpur -- 721302, India\\
$^3$ National Centre for Radio Astrophysics, Tata Institute of
Fundamental Research, Post Bag 3, Ganeshkhind, Pune -- 411007, India\\
$^4$ Department of Physics, Presidency University, 86/1 College
Street, Kolkata -- 700073, India} 

\date{Accepted 2017 November 06. Received 2017 November 06; in original form 2017 June 28}
\pubyear{2017}

\begin{document}
\label{firstpage}
\pagerange{\pageref{firstpage}--\pageref{lastpage}}
\maketitle


\begin{abstract}
The light-cone (LC) effect causes the Epoch of Reionization (EoR)
21-cm signal $T_{\rm b} (\n, \nu)$ to evolve significantly along the
line of sight (LoS) direction $\nu$. In the first part of this paper
we present a method to properly incorporate the LC effect in
simulations of the EoR 21-cm signal that include peculiar
velocities. Subsequently, we discuss how to quantify the second order
statistics of the EoR 21-cm signal in the presence of the LC
effect. We demonstrate that the 3D power spectrum $P(\k)$ fails to
quantify the entire information because it assumes the signal to be
ergodic and periodic, whereas the LC effect breaks these conditions
along the LoS. Considering a LC simulation centered at redshift $8$
where the mean neutral fraction drops from $0.65$ to $0.35$ across the
box, we find that $P(\k)$ misses out $\sim 40 \%$ of the information
at the two ends of the $17.41 \, {\rm MHz}$ simulation bandwidth. The
multi-frequency angular power spectrum (MAPS) $\cl(\nu_1,\nu_2)$
quantifies the statistical properties of $T_{\rm b} (\n, \nu)$ without
assuming the signal to be ergodic and periodic along the LoS. We
expect this to quantify the entire statistical information of the EoR
21-cm signal. We apply MAPS to our LC simulation and present
preliminary results for the EoR 21-cm signal. 
\end{abstract}



\section{Introduction}
\label{sec:intro}
Observation of the redshifted 21-cm signal from neutral hydrogen
(\ion{H}{I}), one of the most promising tools to probe the epoch of
reionization (EoR), is currently a frontier of modern astrophysics and
cosmology. There is a tremendous effort, all over the globe, to detect
the EoR 21-cm signal either statistically or around bright individual 
objects using ongoing and upcoming radio interferometric experiments e.g. 
{GMRT\footnote{\href{http://www.gmrt.ncra.tifr.res.in}{http://www.gmrt.ncra.tifr.res.in}}} \citep{ghosh12, paciga13}, 
{LOFAR\footnote{\href{http://www.lofar.org}{http://www.lofar.org}}} \citep{haarlem13,yatawatta13},
{MWA\footnote{\href{http://www.haystack.mit.edu/ast/arrays/mwa}{http://www.haystack.mit.edu/ast/arrays/mwa}}} \citep{bowman13,tingay13,dillon14}, 
{PAPER\footnote{\href{http://eor.berkeley.edu}{http://eor.berkeley.edu}}} \citep{parsons14,ali15,jacobs14},
{SKA\footnote{\href{http://www.skatelescope.org}{http://www.skatelescope.org}}}
\citep{mellema13,koopmans15} and 
{HERA\footnote{\href{http://reionization.org}{http://reionization.org}}} \citep{furlanetto09}.

One of the major advantages of the redshifted \HI 21-cm signal is that
it allows one to map the large scale structure of the universe in 3D
with the third axis being the cosmic time (or redshift). However, the
mean as well as the statistical properties of \HI 21-cm signal change
with redshift. This effect, known as the `light-cone' (LC) effect, has
a significant impact on the observable quantities such as on \HI 21-cm
brightness temperature maps, power spectrum, etc. It is thus important
to correctly include this effect to predict the signal and to also
interpret the observations.

The issue was first considered in \citet{barkana06} who analytically
modelled the anisotropies in the two-point correlation function
arising due to the LC effect. A similar approach was later followed in
\citet{zawada14} which used large scale numerical simulations and
studied, in more details, the LC anisotropies in the two point
correlation function. \citet{datta12} first investigated the impact of
the LC effect on the spherically averaged \HI power spectrum which is
one of the primary observables for all the ongoing and upcoming radio
interferometric telescopes mentioned earlier. They find that the
effect mainly `averages out' in the spherically averaged power
spectrum and they report a change of up to $\sim 50 \%$ at the large
scales corresponding to a frequency bandwidth of $\sim 8\,{\rm MHz}$. 
Subsequently, ${\rm Gpc}$ size simulations have been used to
investigate various other issues such as quantifying the LC induced
anisotropies in the power spectrum and determining the optimal
bandwidth for analyzing the observed signal in order to avoid
complexities arising from the LC effect \citep{la-plante14,datta14}. 
In a recent work \citealt{ghara15} have considered the \HI 21-cm
signal from the cosmic dawn which includes fluctuations in the spin
temperature. They find that the LC effect has a dramatic signature on
the cosmic dawn \HI power spectrum.

The redshift space distortion due to peculiar velocities is an
important effect that modifies the redshifted 21-cm signal
\citep{bharadwaj04} along the line of sight (LoS). While there has
been considerable work on including this effect in simulations of the
EoR 21-cm signal \citep{mao12,majumdar13, jensen13}, the issue of how
to properly include the LC effect in the presence of peculiar
velocities has not been addressed earlier.

The issue of how to analyze the statistics of the EoR 21-cm signal in
the presence of the LC effect is also important. Note that the signal
in the different Fourier modes is uncorrelated only for a
statistically homogeneous or ergodic signal, and in this case the
second order statistics is completely quantified by the 3D power
spectrum $P(\k)$. However, the LC effect breaks statistical
homogeneity and makes the signal non-ergodic along the LoS. In this
case the signal in the different Fourier modes along the LoS is
correlated. This implies that $P(\k)$ does not retain the entire
information of the 21-cm signal. \citet{trott16} has argued that the
spherically averaged \HI 21-cm power spectrum gives a biased estimate
of the EoR 21-cm signal and has proposed the use of the wavelet
transform to obtain an improved estimate in comparison to the standard
Fourier transform.

In this work we address two issues. First, how to properly incorporate
the LC effect in simulations of the EoR 21-cm signal in the presence of
peculiar velocities. Second, how to properly quantify the statistical 
properties of the EoR 21-cm signal. To this end we consider the
multi-frequency angular power spectrum (MAPS, \citealt{datta07a})
which doesn't assume  the signal to be ergodic along the LoS and
retains the full information of the 21-cm signal.

Throughout this paper, we have used the Planck+WP best fit values of
cosmological parameters $\Omega_{\rm m0}=0.3183$,
$\Omega_{\rm \Lambda0}=0.6817$, $\Omega_{\rm b0}h^2=0.022032$, $h=0.6704$,
$\sigma_8=0.8347$, and $n_{\rm s}=0.9619$ \citep{planck14}.


\section{Simulating the light-cone effect}
\label{sec:LCsim}
The redshifted EoR \HI 21-cm signal is the quantity of interest
here. The Hydrogen distribution evolves dramatically across the
EoR. Starting from the early stages of EoR when the mean mass weighted
Hydrogen neutral fraction $\xb$ is close to $1$, the Hydrogen
distribution evolves rapidly to a situation where it is nearly
completely ionized with $\xb \sim 0$ at the end of reionization. The
issue here is `How to incorporate the light-cone (LC) effect in
simulations of the EoR 21-cm signal?'.

The light-cone (LC) effect refers to the fact that our view of the
Universe is restricted to the backward light cone which imposes the
relation 
\begin{equation}
r = c(\eta_0-\eta) \, ,
\label{eq:lc}
\end{equation}  
between the comoving distance $r$ as measured from our position and
the conformal time $\eta$, the suffix `$0$' here refers to the present
epoch. We consider a simulation that span the comoving distance range
$r_{\rm n}$~(nearest) to $r_{\rm f}$~(farthest). The LC effect implies
that our view at $r_{\rm f}$ is restricted to an early epoch
$\eta_{\rm f}$ (eq.~\ref{eq:lc}) when the universe is largely neutral
whereas at $r_{\rm n}$ it is restricted to a later epoch $\eta_{\rm n}$ 
when the universe is nearly completely reionized. At each distance in
the range $r_{\rm n} \le r \le r_{\rm f}$, we view a different stage
of the cosmological evolution $\eta_{\rm f} \le \eta \le \eta_{\rm n}$
and consequently $\xb$ evolves along the radial direction of the
simulation volume. It is particularly important to account for this
evolution when simulating the EoR 21-cm signal.

\begin{figure}
\psfrag{etai-1}[c][c][1][0]{$\eta_{\rm i-1}$~~}
\psfrag{etai}[c][c][1][0]{$\eta_{\rm i}$~~~}
\psfrag{Coeval}[c][c][1][0]{\large Coeval\,}
\psfrag{ri-1  ri}[c][c][1][0]{~$r_{\rm i-1}\,r_{\rm i}$}
\psfrag{ri  ri+1}[c][c][1][0]{~~$r_{\rm i}\,r_{\rm i+1}$}
\psfrag{Light-cone}[c][c][1][0]{\large Light-cone}
\psfrag{etai-1  eti}[c][c][1][0]{$\eta_{\rm i-1}\,\eta_{\rm i}$~~}
\psfrag{ri-1  ri  ri+1}[c][c][1][0]{~$r_{\rm i-1}\,r_{\rm i}\,r_{\rm i+1}$}
\centering
\includegraphics[width=0.41\textwidth]{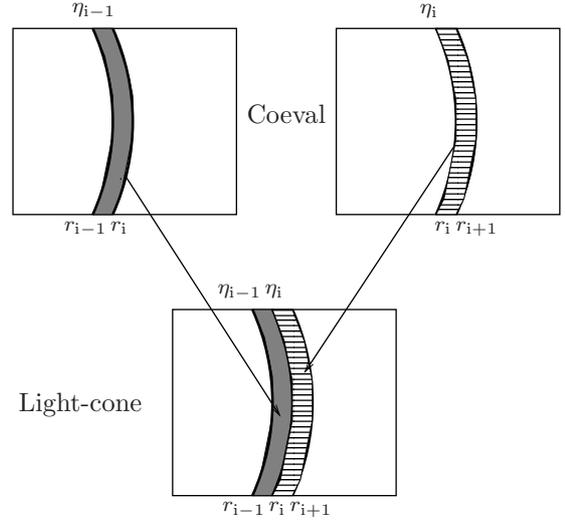}
\caption{This schematically shows how we sliced the coeval simulations
  and stitched the slices to form the LC simulation. The top panels
  represent our coeval simulations (at $\eta_{\rm i-1}$ and $\eta_{\rm i}$ 
respectively) and the bottom panel represents the LC simulation. Note
that the spherical coordinate system has origin at  a distant observer
located on the left.} 
\label{fig:lightcone}
\end{figure}

For our purpose  we have simulated snapshots of the \HI distribution
(so called {\it coeval} cubes) at 25 different epochs $\eta_{\rm i}$
that span the relevant range $\eta_{\rm f} \le \eta_{\rm i} \leq
\eta_{\rm n}$ at non-uniform intervals $\Delta \eta_{\rm i}$ which
were chosen so that $\xb$ varies by approximately an equal amount in
each interval. The \HI distribution in our simulations is represented
by particles whose \HI masses vary with position depending on the
local Hydrogen neutral fraction. Each snapshot provides the positions,
peculiar velocities and \HI masses of these particles. Each epoch
$\eta_{\rm i}$ corresponds to a different radial distance $r_{\rm i}$
in the simulation volume (eq.~\ref{eq:lc}). To construct the LC
simulation  we have sliced the simulation volume at these $r_{\rm i}$,
and for each slice we have filled the region $r_{\rm i}$ to 
$r_{\rm i+1}$ with the \HI particles from the corresponding region in
the snapshot at the epoch $\eta_{\rm i}$ (Fig.~\ref{fig:lightcone}).   


Observations will yield brightness temperature fluctuations $\delta
\Tb(\n,\nu)$ which are measured as a function of the observing
frequency $\nu$ and direction $\n$, here $\n$  is the unit vector in
the direction of observation. For the 21-cm signal originating from
the point $\n\,r$, the cosmological expansion and the radial component
of the \HI peculiar velocity $\n \cdot \v(\n r,\eta)$ together
determine the frequency $\nu$ at which the signal observed, and we
have  
\begin{equation}
\nu=a(\eta) [1- \n \cdot \v(\n r,\eta)/c] \times \nu_{\rm e} \,,
\label{eq:pv}
\end{equation}
where $\nu_{\rm e}=1,420 \, {\rm MHz}$.  For a fixed direction $\n$,
we can view eqs.~(\ref{eq:lc}) and (\ref{eq:pv}) together as a map 
$r \rightarrow \nu$ from comoving distance $r$ to frequency
$\nu$. Considering the different shells within the LC simulation
(Fig.~\ref{fig:lightcone}), we can assign a frequency 
$\nu_{\rm i}=a(\eta_{\rm i}) \nu_{\rm e}$ to the boundary $r_{\rm i}$
of each of these shells. We now consider a simulation  particle
labelled $m$ located at the position $\r_{\rm m} = r_{\rm m} \n_{\rm m}$ 
within the $i$-th shell $(r_{\rm i} \le r_{\rm m} < r_{\rm i+1})$. We
use eq.~(\ref{eq:pv}) to assign the frequency 
\begin{equation}
\nu_{\rm m}=\nu_{\rm i} \left[1 - \frac{a_{\rm i} H_{\rm i} (r_{\rm m} - r_{\rm i})
   + \n_{\rm m} \cdot \v_{\rm m}}{c} \right] 
\label{eq:nu}
\end{equation}
to the redshifted 21-cm signal from the \HI associated with this
particle. Here scale factor $a_{\rm i} \equiv a(\eta_{\rm i})$ and
Hubble parameter $H_{\rm i} \equiv H(\eta_{\rm i})$ respectively. We
use eq.~(\ref{eq:nu}) to map the \HI distribution within the LC
simulation from $\r= r \n$ to  $\nu$ and $\n$ which are the variables
relevant for observations of the 21-cm brightness temperature.

Assuming that the spin temperature is much greater than the background
CMB temperature i.e. $T_{\rm s} \gg T_{\gamma}$, the \HI 21-cm
brightness temperature (eq.~4 and A5 of \citealt{bharadwaj05}) can be
expressed as  
\begin{equation}
T_{\rm b} (\n, \nu) = \bar{T}_0 \, \frac{\rho_{\Hi}}{\bar{\rho}_{\rm H}}\,
\left( \frac{H_{0} \nu_{\rm e}}{c} \right)  \, \bigg| \frac{\partial
  r}{\partial \nu} \bigg| \,, 
\label{eq:bt}
\end{equation}
where
\begin{equation}
\bar{T}_0 = 4.0 {\rm mK}\, \left( \frac{\Omega_{\rm b} h^2}{0.02}\right)
\left( \frac{0.7}{h} \right) \,,
\end{equation}
${\rho_{\Hi}}/{\bar{\rho}_{\rm H}}$ is the ratio of the neutral
hydrogen to the mean hydrogen density, and here it is convenient to
use the respective comoving densities. Eqs.~(\ref{eq:lc}) and
(\ref{eq:pv}) together imply a map from $r$ to $\nu$, and 
${\partial r}/{\partial \nu}$ refers to the derivative of this map.

We note that the comoving \HI density can be obtained by assigning the
\HI mass in the particles to an uniform rectangular grid in comoving
space 
\begin{equation}
\rho_{\Hi}=(\Delta r)^{-3} \sum_m [M_{\Hi}]_{\rm m} 
\label{eq:ro}
\end{equation}
where $(\Delta r)^3$ is the volume of each grid cell. Here we use an
uniform grid in solid angle $(\Delta \Omega)$ and frequency $(\Delta \nu)$ 
to define a modified density $\rho^{\prime}_{\Hi}$ calculated using  
\begin{equation}
\rho^{\prime}_{\Hi}=(\Delta \Omega \, \Delta \nu)^{-1} \left( \frac{H_{0}
  \nu_{\rm e}}{c} \right) \sum_m \frac{[M_{\Hi}]_{\rm m}}{r_{\rm n}^2} \,.
\label{eq:rop}
\end{equation}
Comparing eqs.~(\ref{eq:ro}) and (\ref{eq:rop}) we see that we can
write the brightness temperature (eq. \ref{eq:bt}) in terms of
$\rho^{\prime}_{\Hi}$ as
\begin{equation}
T_{\rm b} (\n, \nu) = \bar{T}_0 \, \frac{\rho^{\prime}_{\Hi}}{\bar{\rho}_{\rm H}}\,.
\label{eq:btp}
\end{equation}
We have used eq.~(\ref{eq:btp}) to calculate the redshifted 21-cm 
brightness temperature distribution $\Tb(\n r)$ from the \HI
distribution in the LC simulation.


\subsection{Generating the coeval cubes}
We have simulated the coeval ionization cubes with a co-moving length
$L=300.16~{\rm Mpc}$ on each side using semi-numerical simulations
which involve three main steps. First, we use a particle mesh $N$-body
code to generate the dark matter distribution. We have run simulations
with $4288^3$ grids of spacing $0.07\,{\rm Mpc}$ and a mass resolution
of $1.09 \times 10^8\,M_{\sun}$. In the next step, we use the
Friends-of-Friends (FoF) algorithm to identify collapsed halos in the
dark matter distribution. We have used a fixed linking length of $0.2$
times the mean inter-particle distance and also set the criterion that
a halo should have at least $10$ dark matter particles. The third and
final step generates the ionization map using the parameters 
$\{N_{\rm ion},\,M_{\rm halo, min},\,R_{\rm mfp}\} =
\{23.21,\, 1.09\times 10^9~M_{\sun},\, 20~{\rm Mpc}\}$ 
(same as \citealt{mondal17,mondal16,mondal15}) based on the excursion
set formalism of \cite{furlanetto04a}. Our semi-numerical simulations
closely follow the homogeneous recombination scheme of
\cite{choudhury09b}. The \HI distribution in our simulations is
represented by particles whose \HI masses were calculated from the
neutral Hydrogen fraction $\xHI$ interpolated from its eight nearest
neighbouring grid points. Each coeval cube provides the positions,
peculiar velocities and \HI masses of these particles. 

\begin{figure}
\psfrag{xh1}[c][c][1][0]{\large $\xb$}
\psfrag{z}[c][c][1][0]{\large Redshift}
\psfrag{r}[c][c][1][0]{\large $r-r_{\rm c}~$Mpc}
\psfrag{10}[c][c][1][0]{\large $10$}
\centering
\includegraphics[width=.45\textwidth]{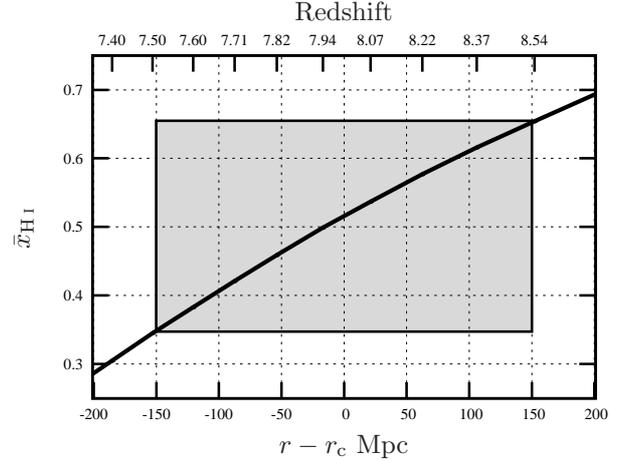}
\caption{This shows the reionization history as a function of
  co-moving distance ($r-r_{\rm c}$) that we have obtained from
  our semi-numerical simulations. We have generated the coeval cubes
  at different $r_{\rm i}$ that span the shaded region. The redshift
  values corresponding to different co-moving distances are shown on
  the top $x$-axis.}
\label{fig:history}
\end{figure}

We have used the semi-numerical simulations to generate the
reionization history, which is shown in Fig.~\ref{fig:history}. We
have generated the coeval cubes at different co-moving distance
$r_{\rm i}$ that span the range $r_{\rm n}=9001.45~{\rm Mpc}$
(nearest) to $r_{\rm f}=9301.61~{\rm Mpc}$ (farthest), which
correspond to the redshifts $7.51$ and $8.53$ respectively. The change
in the mass-averaged \HI fraction $\xb$ over the aforesaid $r$ range,
according to our reionization history (Fig.~\ref{fig:history}), is
$\Delta \xb \approx 0.65-0.35=0.30$. We have chosen $25$ different
$r_{\rm i}$ so that $\xb$ varies by approximately an equal amount in
each interval. Using these coeval \HI cubes, we have generated our
light-cone (LC) box following the formalism presented in
Section~\ref{sec:LCsim}. The LC box is centered at redshift $8$ which
correspond to the co-moving distance $r_{\rm c} = 9151.53~{\rm Mpc}$, 
frequency $\nu_{\rm c}=157.78 \, {\rm MHz}$ and $\xb \approx 0.51$.

\subsection{Flat-sky approximation}
\label{sec:flatsky}
\begin{figure}
\psfrag{nm}[c][c][1][0]{$\n_{\rm m}$}
\psfrag{rm}[c][c][1][0]{$r_{\rm m}$}
\psfrag{theta}[c][c][1][0]{~$\theta_{\rm m}$}
\psfrag{z}[c][c][1][0]{$z$}
\psfrag{o}[c][c][1][0]{~~o}
\psfrag{ri  ri+1}[c][c][1][0]{~~~~$r_{\rm i}\,\,~~~~r_{\rm i+1}$}
\centering
\includegraphics[width=0.45\textwidth]{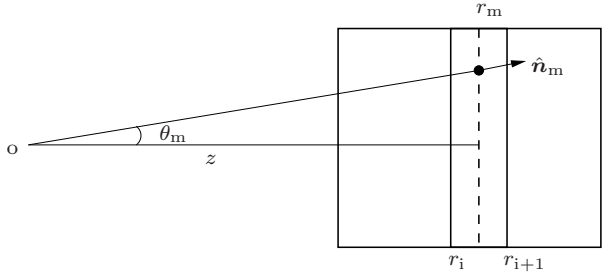}
\caption{This shows the $m$-th particle within $i$-th LC slice at
  $r_{\rm m}$ in the flat-sky approximation.}
\label{fig:flat_sky}
\end{figure}

The observed sky is spherical and the slices simulated at fixed values
of $r_{\rm i}$ are, in general, curved as shown in
Fig.~\ref{fig:lightcone}. However, the angular extent 
$\theta_{\rm  max} =L/(2 r_{\rm f})$ of our simulation box is $\approx
1^{\circ}$ for which it is adequate to adopt the flat-sky
approximation whereby the simulation slices are flat as shown in
Fig.~\ref{fig:flat_sky}. We use a Cartesian coordinate system with the
origin located at the distant observer, the $z$ axis is aligned along
the LoS through the centre of the box, and the $x$ and $y$ axes are in
the plane of the sky -- perpendicular to the $z$-axis. Under the
flat-sky approximation, the unit vector $\n$ along any arbitrary
direction can be decomposed as $\n= \hat{\bm{k}} + \thetavec$ where
$\hat{\bm{k}}$ is the unit vector along the $z$-axis and $\thetavec$
is a 2D vector in the  plane of the sky. The curvature of the sky
introduces terms of order $\theta^2$ and higher which we have ignored
here in the flat-sky approximation. We have, however, retained terms
of order $\theta$ ensuring that the resulting errors are of order 
$< 1\%$. In particular we  use the approximations 
$r=\sqrt{z^2+x^2+y^2} \approx z$,  
$\thetavec \approx [(x/z){\hat{\bm{i}}}+(y/z){\hat{\bm{j}}}]$ and 
$\n_{\rm m} \cdot \v_{\rm  m}  \approx [v_{\rm z}]_{\rm
  m}+(x/z)\,[v_{\rm x}]_{\rm m} + (y/z)\,[v_{\rm y}]_{\rm m}$. 


\begin{figure}
\centering
\includegraphics[width=0.45\textwidth, angle=0]{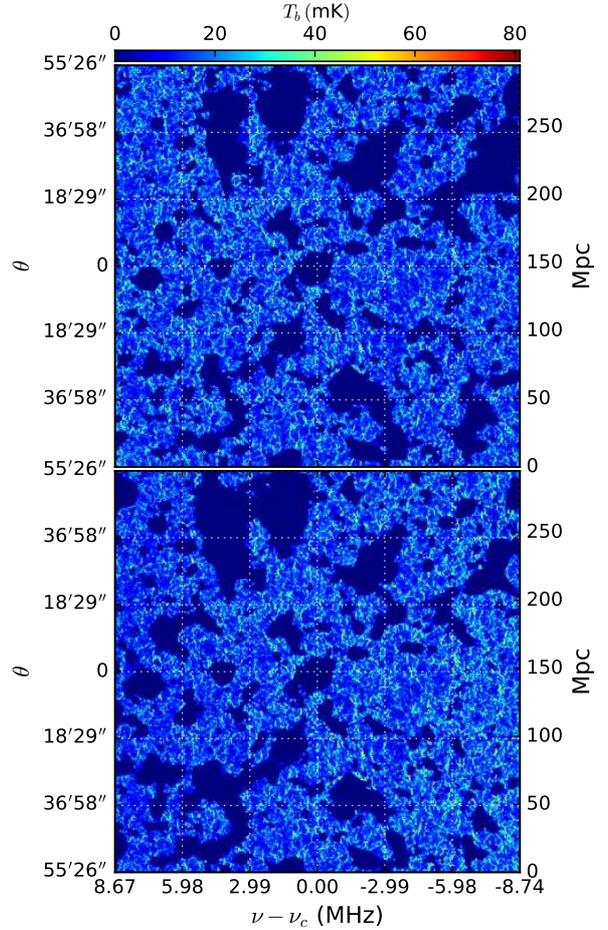}
\caption{This shows $\Tb(\thetavec,\nu)$ on a section through the 3D
  21-cm brightness temperature maps for the LC (bottom) and coeval
  (top) simulations. The right vertical axis and the overlaid grid 
  shows the corresponding comoving coordinates calculated using
  eq.~(\ref{eq:coord}).}
\label{fig:lc_map}
\end{figure}

We use eq.~(\ref{eq:nu}) to map the positions of the \HI particles to
frequency space. The final LC simulation extends from $\nu_{\rm f}$ to
$\nu_{\rm n}$ in frequency, and we note that the inclusion of peculiar
velocities causes some of the particles to have frequency values
beyond the box boundaries. This causes a depletion in the particle
density near the box boundaries. We have  estimated the frequency
interval that is affected by this particle depletion, and we have
excluded slices of this size from both the nearest and farthest sides
of the LC box. Finally, we have interpolated the \HI distribution from
the particles to a 3D rectangular grid in $(\thetavec,\,\nu)$. The
bottom panel of Fig.~\ref{fig:lc_map} shows a section through the
simulated 3D LC 21-cm brightness temperature map. The smaller
frequencies on the right side of the LC simulation correspond to the
earlier stages of the evolution as compared to the larger frequencies
shown on the left side. For comparison, the top panel of
Fig.~\ref{fig:lc_map} shows the same section through a coeval
simulation at the central redshift $8$. The different frequencies in
the coeval simulation all correspond to the same stage of the
evolution. We see that it is possible to identify the same ionized
regions in both the LC and coeval simulations. We see that at the
right side (early stage) the ionized regions appear smaller in the LC
simulation as compared to the coeval case, whereas the ionized regions
appear larger in the LC simulation at the left side (later stage). The
fact that each frequency corresponds to a different stage of the
evolution is clearly evident if we compare the two panels of
Fig.~\ref{fig:lc_map}. We note that the brightness temperature
fluctuations $\delta \Tb(\thetavec,\nu)=T_{\rm b}(\thetavec,\nu) -
\bar{T}_{\rm b}(\nu)$ in the coeval simulations are, by construction,
statistically homogeneous along the LoS  direction $\nu$. The
cosmological evolution seen  in the LC simulation, however, breaks the
statistical homogeneity along the LoS direction $\nu$. The fluctuations 
$\delta \Tb(\thetavec,\nu)$ continue to be statistically homogeneous
along $\thetavec$ in both the coeval and LC simulations. 


\section{Statistical analysis}
The issue here is ``How to quantify the statistics of $\delta
\Tb(\thetavec,\,\nu)$?''.  We consider two statistical quantities
namely the spherically averaged three dimensional (3D) power spectrum
and the multi-frequency angular power spectrum (MAPS) which are
discussed in the two subsequent sub-sections.

\subsection{The power spectrum}
Several authors (see Section~\ref{sec:intro}) have used the 3D power spectrum
$P(\k)$ to quantify the simulated EoR 21-cm signal in the presence  of
the LC effect. The first step here is to map the EoR 21-cm brightness
temperature fluctuations to spatial comoving coordinates 
$\delta T_{\rm b} (\thetavec, \nu) \rightarrow \delta T_{\rm b} (x,y,z)$
within the simulation volume $V$. The fact that $r$ varies with $\nu$
along the LoS and the two have a non-linear relation results in a
spatial grid of non-uniform spacing which poses a problem for
evaluating the Fourier transform needed to compute $P(\k)$. We have
avoided this complication by using 
\begin{equation}
(x,\,y,\,z)=(r_{\rm c}\,\theta_{\rm x},\,r_{\rm c}\,\theta_{\rm y},\,z_{\rm c}
+ r^{\prime}_{\rm c}\,(\nu-\nu_{\rm c})\,)
\label{eq:coord}
\end{equation}
where $r_{\rm c}$ and $r^{\prime}_{\rm c}=\frac{d\,r}{d\,\nu}\big|_{r_{\rm c}}$
are both evaluated at the central redshift of $8$. This approximation
results in a rectangular spatial grid of uniform spacing where we
directly use FFT to estimate $\tTb(\k)$ which is the 3D Fourier
transform of $\delta T_{\rm b}(x,y,z)$. This approximation introduces
an error which is less than $\sim 2  \%$ in grid positions.

\begin{figure}
\psfrag{pk}[c][c][1][0]{\large ${\Delta^2_{\rm b}}\, \, {\rm mK^2}$}
\psfrag{diff}[c][c][1][0]{${\delta \Delta^2_{\rm b}}/\Delta^2_{\rm b}$}
\psfrag{k}[c][c][1][0]{\large $k\,~{\rm Mpc}^{-1}$}
\psfrag{RS}[c][c][1][0]{\large Coeval~~~~}
\psfrag{LC+RS}[c][c][1][0]{\large ~~~~~LC}
\centering
\includegraphics[width=0.45\textwidth, angle=0]{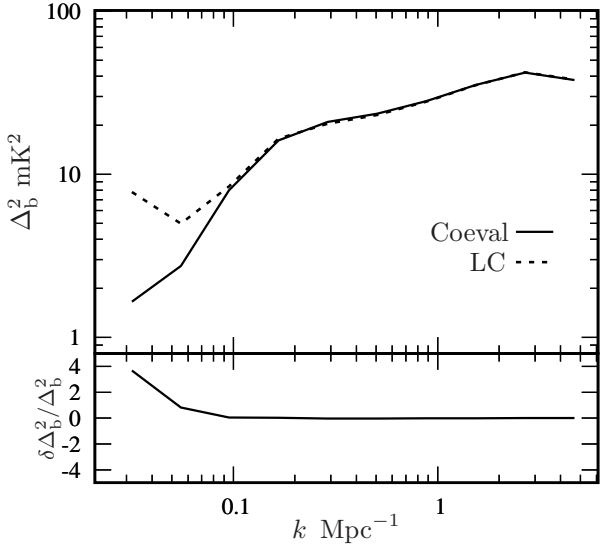}
\caption{This (top panel) shows the dimensionless spherically averaged
  \HI 21-cm power spectrum $\Delta^2_{\rm b} (k)$. The bottom panel
  shows the relative difference ${\delta \Delta^2_{\rm
      b}}/\Delta^2_{\rm b} = \frac{[\Delta^2_{\rm b}]_{\rm
      LC}-[\Delta^2_{\rm b}]_{\rm Coeval}}{[\Delta^2_{\rm b}]_{\rm
      Coeval}}$.}  
\label{fig:pk}
\end{figure}

The 3D \HI 21-cm power spectrum can be calculated using
\begin{equation}
P(\k) = V^{-1}\big\langle \tTb(\k)\,\tTb(-\k)\big\rangle\,.
\label{eq:pk}
\end{equation}
Fig.~\ref{fig:pk} shows the  dimensionless spherically averaged \HI
21-cm power spectra $\Delta^2_{\rm b} (k) = k^3 P(k)/2\pi^2$ as a
function of $k$ for the LC and coeval simulations, both centred at redshift
$8$. We see that the LC effect introduces a very significant
enhancement at large scales and the two power spectra differ by factors 
of $\sim 4$ and $2$ at $k \sim 0.03\,{\rm Mpc}^{-1}$ and $0.05\,{\rm
  Mpc}^{-1}$ respectively. We note that these large scale modes are
affected by the sample variance due to the finite size of simulation
cubes used and actual value might change to some extent. Although the
simulation methodology and the parameters used here are quite
different, this result is consistent and qualitatively similar to
those obtained earlier \citep{datta12,la-plante14,datta14}.


It is important to note that the EoR 21-cm signal $\delta T_{\rm b}
(\thetavec, \nu)$ evolves significantly along the LoS direction $\nu$
due to the LC effect (Fig.~\ref{fig:lc_map}). While the 3D Fourier
modes and 3D power spectrum $P(\k)$ are optimal if the signal is
statistically homogeneous, the 3D Fourier modes which are used to
calculate $P(\k)$ are not the optimal basis set when the statistical
properties of the signal evolve within the simulation
volume. Additionally, the Fourier transform imposes periodicity on the
signal, an assumption that cannot be justified along the LoS once the
LC effect is included. These effects imply that the 3D power spectrum
fails to fully quantify the entire signal. These effects can also
introduce artefacts in the 3D power spectrum estimation
\citep{trott16}.   


\subsection{The multi-frequency angular power spectrum}
Here we decompose the brightness temperature fluctuations
$\delta T_{\rm b} (\n,\,\nu)$ in terms of spherical harmonics
$Y_{\ell}^{\rm m}(\n)$ using 
\begin{equation}
\delta T_{\rm b} (\n,\,\nu)=\sum_{\ell,m} a_{\ell {\rm m}} (\nu) \,
Y_{\ell}^{\rm m}(\n)
\label{eq:alm}
\end{equation}
and define the multi-frequency angular power spectrum (hereafter MAPS,
\citealt{datta07a}) as
\begin{equation}
\cl(\nu_1, \nu_2) = \big\langle a_{\ell {\rm m}} (\nu_1)\, a^*_{\ell
  {\rm m}} (\nu_2) \big\rangle\, .
\label{eq:cl}
\end{equation}
This incorporates the assumption that the EoR 21-cm signal is
statistically homogeneous and isotropic with respect to different
directions in the sky, however the signal is not assumed to be
statistically homogeneous along the LoS direction $\nu$. We expect
$\cl(\nu_1,\nu_2)$ to entirely quantify the second order statistics
of the EoR 21-cm signal.

In the present work it suffices to adopt the flat sky approximation
where we decompose the $\thetavec$ dependence of $\delta T_{\rm b}
(\thetavec,\nu)$ into 2D Fourier modes $\TTb(\U, \nu)$. Here
$\U$ is the 
Fourier conjugate of $\thetavec$, and we define the MAPS using 
\begin{equation}
\cl(\nu_1,\,\nu_2) = {\mathcal C}_{2\pi{\rm U}}(\nu_1,\,\nu_2) =
\Omega^{-1}\,\big\langle \TTb(\U,\,\nu_1)\,
\TTb(-\U,\,\nu_2)\big\rangle\,
\label{eq:cl_flat}
\end{equation}
where $\Omega$ is the solid angle subtended by the simulation at the
observer. 

\begin{figure*}
\psfrag{cl}[c][c][1][0]{$\ell(\ell+1)\, \cl(\nu_1, \nu_2)/(2\pi)~{\rm mK}^2$}
\psfrag{nu1}[c][c][1][0]{$\nu_1 - \nu_{\rm c}$~MHz}
\psfrag{nu2}[c][c][1][0]{$\nu_2 - \nu_{\rm c}$~MHz}
\centering
\includegraphics[width=0.99\textwidth, angle=0]{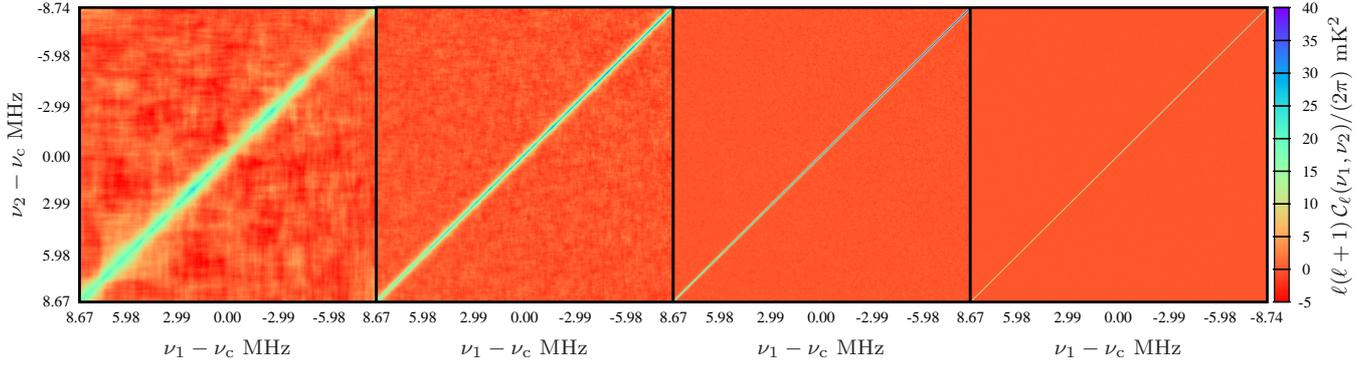}
\caption{This shows the multi-frequency angular power spectrum
  $\ell(\ell+1)\, \cl(\nu_1, \nu_2)/(2\pi)$ at $\ell=1468,\,4486,\,13728$
  and $42018$ (left to right panels) for the LC simulation.} 
\label{fig:cl_nu1nu2}
\end{figure*}

\begin{figure}
\psfrag{cl}[c][c][1][0]{\large $\ell(\ell+1)\cl^{\rm EP}(\Delta
  \nu)/(2\pi)~{\rm mK}^2$}  
\psfrag{deltanu}[c][c][1][0]{$\Delta \nu$~{MHz}}
\psfrag{l=1468}[c][c][1][0]{$\ell=1468$}
\psfrag{4486}[c][c][1][0]{~~\,$4486$}
\psfrag{13728}[c][c][1][0]{~~\,$13728$}
\psfrag{42018}[c][c][1][0]{~~\,$42018$}
\centering
\includegraphics[width=0.45\textwidth, angle=0]{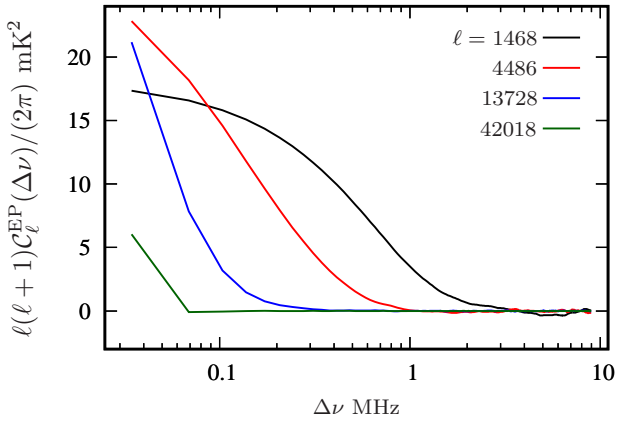}
\caption{This shows $\ell(\ell+1)\, \cl^{\rm EP}(\Delta \nu)/ 2 \pi$
  as a function of $\Delta \nu$ for the LC simulation for the four
  different  $\ell$ bins considered in Figure~\ref{fig:cl_nu1nu2}. The
  $\Delta \nu$ values have been shown for only half the bandwidth as
  the signal is periodic.}
\label{fig:cl_deltanu}
\end{figure}


The $\ell$ range $\ell_{\rm min}=2\pi/\theta_{\rm max}=195$ to
  $\ell_{\rm max}=2\pi/\theta_{\rm min}=52178$ corresponding to our
LC simulation was divided in 10 equally spaced logarithmic bins,
and we have computed the average $\cl(\nu_1, \nu_2)$ for each of
these bins. Fig.~\ref{fig:cl_nu1nu2} shows $\ell(\ell+1)\,
\cl(\nu_1-\nu_c,\,\nu_2-\nu_c)/ 2 \pi$ estimated from our LC
simulation at $\ell=1468,\,4486,\,13728$ and $42018$, where
$\nu_c=157.78 \, {\rm MHz}$ corresponding to the central redshift
$z=8$ of the LC simulations. We see that the signal peaks along the
diagonal elements $\nu_1=\nu_2$ of $\cl(\nu_1,\nu_2)$ and falls
rapidly away from the diagonal i.e. as the frequency separation 
$\Delta \nu =\,\,\mid\nu_1 - \nu_2\mid$ is increased. It is more clear
in Fig.~\ref{fig:cl_deltanu} where we see that MAPS falls by atleast
an order of magnitude beyond $\Delta \nu= 0.5\,{\rm MHz}$ for
$\ell=4486$. It oscillates close to zero with both positive and
  negative $\cl$ values for even larger $\Delta \nu$. The behaviour is 
similar for the other multipoles. The value of $\cl(\nu_1, \nu_2)$ also falls off more
rapidly away from the diagonal as the value of $\ell$ is increased. We
do not discuss these features in any further detail here, and plan to
present this in future work.  

\begin{figure}
\psfrag{cl}[c][c][1][0]{\large $\ell(\ell+1)\cl(\nu, \nu)/(2\pi)\,~\rm
  mK^2$\hspace{2.9cm}} 
\psfrag{nu}[c][c][1][0]{$\nu - \nu_{\rm c}$\,~MHz}
\psfrag{EP}[c][c][1][0]{\large EP}
\psfrag{LC}[c][c][1][0]{\large LC}
\psfrag{l=1468}[c][c][1][0]{$\ell=1468$}
\psfrag{l=4486}[c][c][1][0]{$\ell=4486$}
\psfrag{l=13728}[c][c][1][0]{$\ell=13728$}
\psfrag{l=42018}[c][c][1][0]{$\ell=42018$}
\centering
\includegraphics[width=0.45\textwidth, angle=0]{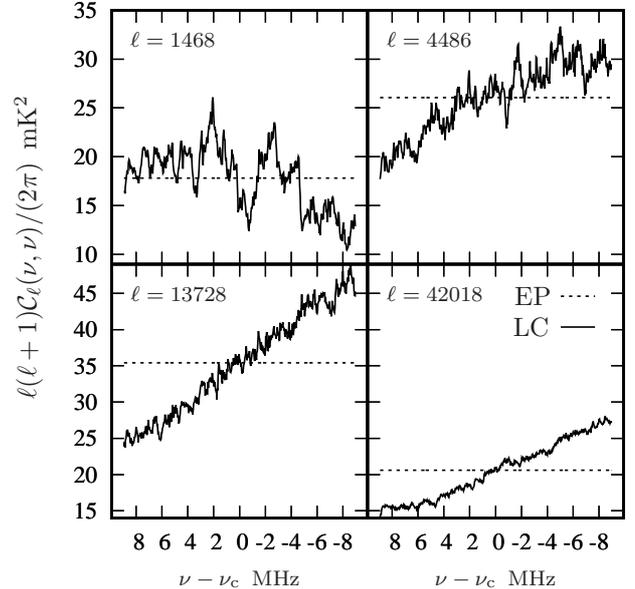}
\caption{This shows $\ell(\ell+1)\, \cl(\nu, \nu)/2 \pi$ as a function
  of $\nu$ for the LC simulation at the four different $\ell$ bins
  considered in Fig.~\ref{fig:cl_nu1nu2}. The values of
  $\ell(\ell+1)\, \cl^{{\rm EP}}(\nu, \nu)/2 \pi$ (horizontal straight
  line) have been shown for comparison.}
\label{fig:cl_lcEP}
\end{figure}
\begin{figure*}
\psfrag{delta-cl}[c][c][1][0]{$\delta \cl/\cl$}
\psfrag{nu}[c][c][1][0]{$\nu - \nu_{\rm c}$\,~MHz}
\psfrag{RS}[c][c][1][0]{\large Coeval~~~~}
\psfrag{LC+RS}[c][c][1][0]{\large ~~~~~LC}
\psfrag{l=1468}[c][c][1][0]{$\ell=1468$}
\psfrag{l=4486}[c][c][1][0]{$\ell=4486$}
\psfrag{l=13728}[c][c][1][0]{$\ell=13728$}
\psfrag{l=42018}[c][c][1][0]{$\ell=42018$}
\centering
\includegraphics[width=1\textwidth, angle=0]{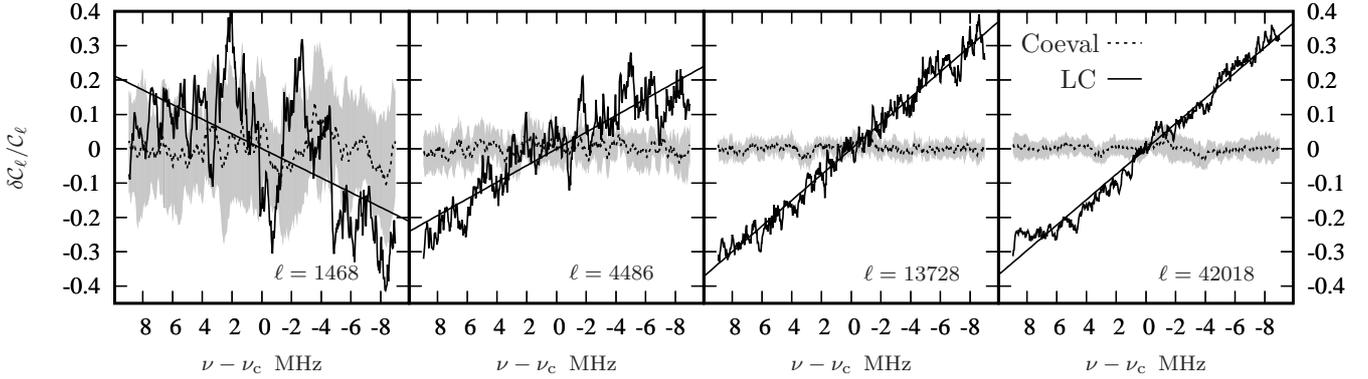}
\caption{The dimensionless $\delta \cl/\cl$ for the $\ell$ values
  shown in the figure. We also show straight line fits to  $\delta
  \cl/\cl$ estimated from the LC simulation. We have used 
$10$ statistically independent realizations of the coeval simulation
to estimate the mean $\delta \cl/\cl$ which is close to zero, and 
the $1-\sigma$ fluctuations  which have been shown by the shaded
regions.}
\label{fig:delta_cl}
\end{figure*}


We now consider the relation between $P(\k)$ and $\cl(\nu_1,\nu_2)$.
As mentioned earlier, $P(\k)$ assumes that the signal is ergodic (E)
and periodic (P) along the LoS direction. We define
$\cl^{{\rm EP}}(\nu_1,\nu_2)$ which is the ergodic and periodic
component of $\cl(\nu_1,\nu_2)$. We estimate $\cl^{ {\rm EP}}
(\nu_1,\nu_2)$ from the measured $\cl(\nu_1,\nu_2)$ by imposing the
conditions $\cl^{{\rm EP}}(\nu_1,\nu_2)=\cl^{{\rm EP}}(\Delta \nu)$
(ergodic) and $\cl^{{\rm EP}}(\Delta \nu) =\cl^{{\rm EP}}(B-\Delta \nu)$
(periodic) where $B$ is the frequency bandwidth of the simulation. 
In the flat sky approximation, $P(\k)$ is the Fourier transform of
$\cl^{{\rm EP}}(\Delta \nu)$, and we have \citep{datta07a}  
\begin{equation}
P(k_{\perp},\,k_{\parallel})= r_{\rm c}^2\,r^{\prime}_{\rm c} \int d (\Delta \nu) \,
  e^{-i  k_{\parallel} r^{\prime}_{\rm c} \Delta  \nu}\, \cl^{\rm EP}(\Delta \nu)
\label{eq:cl_Pk}
\end{equation}
where $k_{\parallel}$ and $k_{\perp}=\ell/r_{\rm c}$ are the
components of $\k$ respectively parallel and perpendicular to the
LoS. A brief derivation of eq.~(\ref{eq:cl_Pk}) is presented in the
Appendix~\ref{a1}. Fig.~\ref{fig:cl_deltanu} shows $\cl^{{\rm
    EP}}(\Delta \nu)$ estimated from our LC simulation. It is
essentially an average of the quantity $\ell(\ell+1)\, \cl(\nu_1,
\nu_2)/ 2 \pi$ over all possible combination of $\nu_1$, $\nu_2$ shown
in Fig.~\ref{fig:cl_nu1nu2} for a given frequency separation $\Delta
\nu=\nu_1-\nu_2$. We also impose the periodicity condition i.e,
$\cl(\Delta \nu) =\cl(B-\Delta \nu)$ while calculating $\cl^{{\rm
    EP}}(\Delta \nu)$. We see that the signal decorrelates rapidly as
$\Delta \nu$ increases, and the decorrelation is more rapid at larger
$\ell$ values consistent with the behaviour seen in
Fig.~\ref{fig:cl_nu1nu2}. This can be understood from eq. A5 which
shows that $\cl^{\rm EP}(\Delta \nu)$ is a Fourier transform of the
\HI power power spectrum $P(k_{\perp}, k_{\parallel})$ along the the
$k_{\parallel}$ axis, where $k_{\perp}=2 \pi \ell/ r_c$. $P(k_{\perp},
k_{\parallel})$ effectively remains flat up to modes $k_{\parallel}
\lesssim k_{\perp}$ when plotted as a function of $k_{\parallel}$. For
large values of $\ell$, the spread of this flatness along
$k_{\parallel}$ gets higher. This results in a steeper Fourier
transform for larger $\ell$ i.e, faster decorrelation of $\cl^{\rm
  EP}(\Delta \nu)$. We have estimated $\cl^{{\rm EP}}(\Delta \nu)$
from our LC simulation, and used this in eq.~(\ref{eq:cl_Pk}) to
calculate $P(\k)$. Fig.~\ref{fig:pk_comp} presents a comparison of
$P(k)$ calculated using eq.~(\ref{eq:cl_Pk}) with that obtained
directly from the 3D Fourier transform (Fig.~\ref{fig:pk}), we find
that the two agree to an accuracy better than $1 \%$. We also note
that the quantity  $\ell(\ell+1)\,  \cl(\Delta \nu )/ 2 \pi$, which
represents the power of fluctuations at scale $\ell$, first increases
and then decreases with $\ell$ when $\Delta \nu$ is very small. This
`peak' in the MAPS corresponds to the characteristic scale of ionized
regions (see \citealt{datta07a} for details).

The MAPS $\cl(\nu_1,\nu_2)$ quantifies the entire second order
statistics of the EoR 21-cm signal even in the presence of the LC
effect. In comparison to this, the 3D power spectrum $P(\k)$ only
quantifies a part of this information, namely the part contained in
$\cl^{{\rm EP}}(\Delta \nu)$. The difference $\delta \cl(\nu_1,\nu_2)
= \cl(\nu_1,\nu_2) - \cl^{{\rm EP}}(\nu_1,\nu_2)$ provides an estimate
of the information that is missed out by the 3D power spectrum
$P(\k)$. Here we focus on the diagonal elements $\nu_1=\nu_2$ where
the MAPS signal peaks
(Fig.~\ref{fig:cl_nu1nu2}). Fig.~\ref{fig:cl_lcEP} shows how the
diagonal element $\ell(\ell+1)\cl(\nu,\nu)/2 \pi$ varies with
$\nu$. We see that $\ell(\ell+1) \cl(\nu,\nu)/2 \pi$ increases with
decreasing $\nu$ which corresponds to increasing neutral fraction
along the LoS direction. For comparison we also show $\ell(\ell+1)
\cl^{{\rm EP}}(\nu,\nu)/ 2 \pi$ which does not vary  with $\nu$.  
Fig.~\ref{fig:delta_cl} shows how $\delta
\cl/\cl^{\rm EP}$ varies with $\nu$ for different values of $\ell$,
note that the denominator here does not vary with $\nu_1$ for the
diagonal terms. For comparison we also show the results for the coeval
simulation centered at redshift $8$. The coeval simulation is ergodic
and has periodic boundary conditions along the LoS, and we expect
$P(\k)$ to work perfectly well in this case. We see that $\delta
\cl/\cl$ estimated from the coeval simulations exhibits random
fluctuations around zero, and is roughly consistent with zero. We
interpret these random fluctuations as arising due to cosmic
variance. The magnitude of these fluctuation become smaller as we go
to larger $\ell$. We can explain this by noting  that the number of
independent $\ell$ modes in each bin increases with $\ell$ for the
logarithmic binning adopted here. In contrast to the coeval
simulation, we find that $\delta \cl/\cl^{\rm EP}$ shows a systematic
variation with $\nu_{\rm i}$ in the LC simulation. This variation is
particularly pronounced at large $\ell$ where the value of $\delta
\cl/\cl^{\rm EP}$ varies systematically from $\sim -0.4$ to $\sim 0.4$
with decreasing frequency across the  bandwidth of our
simulation. This clearly indicates that the 3D power spectrum misses
out $\sim 40 \%$ of the information at the two ends of our $17.41 \,
{\rm MHz}$ band. 

We note that the smallest $\ell$ bin shown in Fig.~\ref{fig:delta_cl}
shows a different behaviour compared to the larger $\ell$
bins. However it is important to note that the smaller $\ell$ bins
also have a larger cosmic variance. 


\section{Discussion and Conclusions}
We first present a method to properly incorporate the LC effect in
simulations of the EoR 21-cm signal in the presence of peculiar
velocities. The method is implemented using a suite of coeval
simulations which we have sliced and stitched together along the LoS
direction to construct  the LC simulation. Our simulation box,
centered at  redshift $8$, subtends $\sim 17.41 \, {\rm MHz}$ along
the LoS and $\xb$ drops from $0.65$ to $0.35$ across the box due to
the LC effect. The statistical properties of the 21-cm signal also
evolve significantly in the LoS direction.

The 3D \HI 21-cm power spectrum $P(\k)$ assumes the signal to be
ergodic and periodic. The LC effect breaks both these properties along
the LoS direction, and as a consequence $P(\k)$ fails to quantify the
entire second order statistics. Here we consider the multi-frequency
angular power spectrum (MAPS) $\cl(\nu_1,\nu_2)$ which  does not
assume the signal to be ergodic and periodic along the LoS. We expect
MAPS to quantify the entire second order statistics of the EoR 21-cm
signal.

We show that it is possible to entirely recover $P(\k)$ from $\cl^{\rm
  EP}(\nu_1,\nu_2)$ which is the ergodic and periodic component of
$\cl(\nu_1,\nu_2)$, and $P(\k)$ misses out the information contained
in $\delta \cl=\cl-\cl^{\rm EP}$. Considering the diagonal elements 
$(\nu_1=\nu_2)$ of $\cl(\nu_1,\nu_2)$, we use the ratio $\delta
\cl/\cl^{\rm EP}$ to quantify the non-ergodicity introduced by the LC 
effect. At small angular scales $\ell \sim 4\times10^3-4\times10^4$ we
find that $\delta \cl/\cl^{\rm  EP}$ shows a systematic increase from
$\sim -0.4$ to $\sim 0.4$ from the largest to the smallest frequency
which respectively correspond to the nearest and furthest ends of the
box along the LoS. This result correlates very well with the fact that
mean neutral fraction increases along the LoS, and we expect
$\cl(\nu_1,\nu_2)$ to increase as we move from the nearest to the
furthest end of the box. The cosmic variance dominates at large
angular scales $\ell \lesssim 10^3$, and we possibly need larger
simulations to address this range.

Our work indicates that $P(\k)$ fails to quantify the entire 21-cm
signal, and we find that it misses out $30 \% - 40 \%$ of the
information at the two end of the  $17.41 \, {\rm MHz}$ frequency band
of our simulation  due to the LC effect. In contrast, we expect MAPS
$\cl(\nu_1,\nu_2)$ to quantify the entire second order statistics of
the EoR 21-cm signal. MAPS is also directly related to the
correlations between the visibilities that are measured in
radio-interferometric observation and it is, in principle, relatively
straightforward to estimate this from observation
\citep{ali08,Ghosh11}. In future work we plan to present more detailed
predictions for the expected EoR 21-cm signal in terms of MAPS.


\section*{Acknowledgements}
RM would like to acknowledge Anjan Kumar Sarkar for his help.


\bibliographystyle{mnras} 
\bibliography{refs}

\appendix
\section{Comparison of power spectra}
\label{a1}
Assuming the 21-cm signal $\delta \Tb(\r)$ to be ergodic and periodic
in the volume $V$, we decompose this into 3D Fourier modes as
\begin{equation}
\delta \Tb(\r)= V^{-1} \sum_{\k}\,{\rm e}^{-i\k\cdot \r}\,\tTb(\k)\,.  
\label{eq:3d}  
\end{equation}
Using eq.~(\ref{eq:coord}), this can be also written as 
\begin{equation}
\delta \Tb(\r) = V^{-1} \sum_{\k}\, {\rm e}^{-i r_{\rm c} \k_{\perp} \cdot
   \thetavec}~{\rm e}^{ - i k_{\parallel} [z_{\rm c}+ r^{\prime}_{\rm
       c}\,(\nu-\nu_{\rm c})]} \,  \tTb(\k_{\perp},k_{\parallel}) \,. 
\label{eq:3d1} 
\end{equation}
The same signal can also be decomposed into 2D Fourier modes as 
\begin{equation}
\delta T_{\rm b}(\thetavec,\nu) = \Omega^{-1} \sum_{\U}
{\rm e}^{-2\pi i   \U \cdot \thetavec} \,  \TTb(\U, \nu)\,. 
\label{eq:2d}  
\end{equation}
Comparing  eq.~(\ref{eq:3d1}) and eq.~(\ref{eq:2d}), we can identify 
$\k_{\perp}=2 \pi \U/r_{\rm c}$ and 
\begin{equation}
\TTb(\U, \nu)  = \Omega V^{-1} \,  \sum_{k_{\parallel}} 
{\rm e}^{ - i k_{\parallel} [z_{\rm c}+ r^{\prime}_{\rm
       c}\,(\nu-\nu_{\rm c})]}  \,  \tTb(\k_{\perp},k_{\parallel}) \,.
\label{eq:2da}  
\end{equation}
We use this in eq.~(\ref{eq:cl_flat}) to calculate $\cl^{\rm EP}(\Delta
\nu) \equiv \cl^{\rm EP}(\nu,\nu+\Delta \nu)$ with $\ell=2 \pi \mid \U
\mid$. This gives 
\begin{equation}
\cl^{\rm EP}(\Delta\nu)=(r_{\rm c}^2 r^{\prime}_{\rm c} B)^{-1}
\sum_{k_{\parallel}} {\rm e}^{i k_{\parallel}  r^{\prime}_{\rm c} \Delta\nu}~
P(k_{\perp},k_{\parallel})\,,
\label{eq:pkcl}
\end{equation}
where we have used the fact that $V=r_{\rm c}^2 r^{\prime}_{\rm c}
\Omega \, B$, and 
\begin{equation}
\big \langle \tTb(\k) \tTb(\k^{\prime}) \big\rangle= \delta_{\k,\k^{\prime}}
\, V \, P(\k)
\label{eq:ergpk}
\end{equation}
which holds when the signal is ergodic. Here $\delta_{\k,\k^{\prime}}$
is the Kronecker delta. We obtain eq.~(\ref{eq:cl_Pk}) which allows us
to calculate $P(k_{\perp},\,k_{\parallel})$ in terms of $\cl^{\rm
  EP}(\Delta \nu)$ by inverting the Fourier relation in
eq.~(\ref{eq:pkcl}). 

\begin{figure}
\psfrag{pk}[c][c][1][0]{\large ${\Delta^2_{\rm b}}\,~{\rm mK^2}$}
\psfrag{diff}[c][c][1][0]{$100\times{\delta \Delta^2_{\rm b}}/\Delta^2_{\rm b}$}
\psfrag{k}[c][c][1][0]{\large $k\,~{\rm Mpc}^{-1}$}
\psfrag{0.5*3D}[c][c][1][0]{\large $0.5\times 3$D~~}
\psfrag{EP}[c][c][1][0]{\large ~EP}
\centering
\includegraphics[width=0.45\textwidth, angle=0]{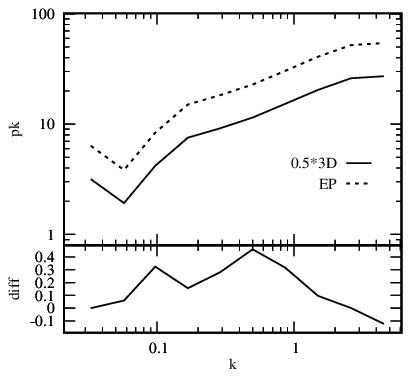}
\caption{This shows the dimensionless spherically averaged \HI 21-cm
  power spectrum $\Delta^2_{\rm b} (k)$ estimated using two different
  methods for the LC simulation. . The bottom panel shows the
  relative difference ${\delta \Delta^2_{\rm b}}/\Delta^2_{\rm b} =
  \frac{[\Delta^2_{\rm b}]_{\rm EP}-[\Delta^2_{\rm b}]_{\rm
      3D}}{[\Delta^2_{\rm b}]_{\rm 3D}}$.}
\label{fig:pk_comp}
\end{figure}

Fig.~\ref{fig:pk_comp} shows a comparison of the dimensionless
spherically averaged \HI 21-cm power spectrum $\Delta^2_{\rm b} (k)$
calculated directly using the 3D Fourier modes (eq.~\ref{eq:pk}) and
the same quantity calculated from $\cl^{\rm EP}(\Delta \nu)$ using 
eq.~(\ref{eq:cl_Pk}). We see  that the two methods give results which
agree to a high level of accuracy, the differences being less than $1
\%$. 
\vfill
\bsp

\label{lastpage}

\end{document}